# Universal criterion and phase diagram for switching a magnetic vortex core in soft magnetic nanodots

Ki-Suk Lee,[1] Sang-Koog Kim,[1]* Young-Sang Yu,[1] Youn-Seok Choi,[1] Konstantin Yu. Guslienko,[1]

Hyunsung Jung,[1] and Peter Fischer[2]

[1]*Research Center for Spin Dynamics and Spin-Wave Devices, and Nanospinics Laboratory, Department of Materials Science and Engineering, College of Engineering, Seoul National University, Seoul 151-744, Republic of Korea*

[2]*Center for X-Ray Optics, Lawrence Berkeley National Laboratory, 1 Cyclotron Road, Mail Stop 2R0400, Berkeley, California 94720, USA*

The universal criterion for ultrafast vortex-core switching between the up- and down-core bistates in soft magnetic nanodots was investigated by micromagnetic simulations along with analytical calculations. Vortex-core switching occurs whenever the velocity of vortex-core motion reaches the critical velocity that is expressed as $\upsilon_{\mathrm{cri}} = (1.66 \pm 0.18)\gamma\sqrt{A_{\mathrm{ex}}}$ (e.g. $\upsilon_{\mathrm{cri}} = 330 \pm 37$ m/s for Permalloy), where $A_{\mathrm{ex}}$ is the exchange stiffness, and $\gamma$ is the gyromagnetic ratio. On the basis of the above results, phase diagrams for the vortex-core switching event and switching times with respect to both the amplitude and frequency of applied *circularly rotating* magnetic field were calculated, which offer practical guidance for implementing nanodots in vortex states into future solid-state information-storage devices.

In magnetic thin films [1,2] and patterned magnetic elements of micrometer (or smaller) lateral size [3], a nontrivial, non-uniform magnetization (**M**) configuration has been experimentally observed in both static and dynamic states. This magnetic nanostructure, the so-called "magnetic vortex", has an in-plane curling **M** along with an out-of-plane **M** at the core area stretching over a few tens of nm in size [3]. Owing to a high thermal stability of this static structure as well as the bistate **M** orientations of the tiny vortex core (VC), the magnetic vortex has been received considerable attention as an information carrier of binary digits "0" and "1" in future nonvolatile information-storage devices [4]. Furthermore, very recently, experimental, theoretical, and simulation studies have explored a rich variety of the dynamic properties of the magnetic vortex, including ultrafast VC switching by linearly oscillating [5-7] and circularly rotating [8-10] in-plane magnetic fields or spin-polarized ac currents [11,12], with extremely low power consumption. The underlying mechanism and physical origin of ultrafast VC switching have also been found [5-7,13]. These rich dynamic properties stimulate continuing intensive studies of patterned magnetic elements in vortex states targeting towards a fundamental understanding of their dynamics [6,7,11,13,14] and real applications to a new class of nonvolatile random access memory [15] and patterned information storage media [4,9]. Such new conceptual devices using ultrafast, low-power VC switching becomes an emerging key issue in the research areas of nanomagnetism and **M** dynamics. Although, the fundamental understanding of the VC reversal and vortex gyrotropic dynamics were much advanced recently, the universal criterion for VC switching, its phase diagram, and switching time have not been investigated yet. Moreover, these are technologically essential parameters for a reliable manipulation of the VC switching acting as a basic function in information-storage devices, which should be identified for its practical applications.

In this Letter, we report on the universal criterion, namely the critical velocity $\upsilon_{cri}$ of VC motions required for VC switching, as found by micromagnetic simulations and analytical calculations. On the basis of the universality of $\upsilon_{cri}$ we derive phase diagrams of the VC reversal event and switching times with respect to the amplitude and frequency of *circularly rotating* magnetic field.

In the present study, the OOMMF code [16] was used that utilizes the Landau-Lifshitz-Gilbert (LLG) equation of motion [17] because this approach is a well established, optimized tool, and reliable enough to investigate **M** dynamics on a few nm spatial and 10 ps temporal scales. In addition, we used an analytical approach to determine the threshold of driving forces required for VC switching and necessary switching time, based on the linearized Thiele's equation [18] of motion. We used Permalloy (Py: $Ni_{80}Fe_{20}$) nanodots as a model system, and a constant Gilbert damping parameter $α = 0.01$, each dot with a different radius $R$ ranging from 150 to 600 nm and a different thickness $L$ ranging from 10 to 50 nm [see Fig. 1(a)].

To excite vortex gyrotropic motions up to the VC switching, we used a specially designed driving force of counter-clockwise (CCW) circularly rotating magnetic fields in the dot plane with the angular frequency $\omega_\mathbf{H}$ and the amplitude $H_0$, denoted as $\mathbf{H}_{\text{CCW}} = H_0 \left[ \cos(\omega_\mathbf{H} t)\hat{\mathbf{x}} + \sin(\omega_\mathbf{H} t)\hat{\mathbf{y}} \right]$ [19,20]. The reason for selecting this $\mathbf{H}_{\text{CCW}}$ is that it is most effective for selective resonant excitations of only the core-up vortex state, as demonstrated in earlier theoretical and simulation works [9,20], and in recent experimental work [10]. An example of the resonant vortex gyrotropic motion and VC switching driven by $\mathbf{H}_{\text{CCW}}$ with $H_0$ = 20 Oe and $\nu_\mathbf{H} = \omega_\mathbf{H}/2\pi$ = 580 MHz (where $\nu_D = \omega_D/2\pi$ = 580 MHz is the vortex eigenfrequency) [21,22] is shown in Fig. 1(b). The orbital trajectories of the earlier transient and steady-state motions of the initial up core and its reversed down core, and their velocities following their individual orbital trajectories are revealed. It is numerically found that the up-core switches to the down-core when the velocity of the up-core motion reaches a threshold velocity, $\upsilon_{\text{cri}}^{\text{Py}}$ = 330 m/s for the Py dot, as seen in the right panel of Fig. 1(b), and in our earlier works [9,12,13].

To examine the universality of this value of $\upsilon_{\text{cri}}^{\text{Py}}$ = 330 m/s, we conducted additional simulations to obtain the VC velocity-versus-time curves varying both $H_0$ (10 to 350 Oe) and $\nu_\mathbf{H}$ (0.1 to 2 GHz) for a Py dot of $R$ = 150 nm and $L$ = 20 nm, as shown in Fig. 2(a). It is established that the value of $\upsilon_{\text{cri}}^{\text{Py}}$ = 330 ± 37 m/s is neither affected by the external field parameters of $\nu_\mathbf{H}$ and $H_0$ nor by the size of the Py dot, as evidenced by the independence of $\upsilon_{\text{cri}}$ on the dot radius and thickness [Fig. 2(b)]. Furthermore, in order to examine whether any of the intrinsic material parameters affects the value of $\upsilon_{\text{cri}}^{\text{Py}}$, we performed simulations for a circular dot of $R = 150$ nm and $L = 20$ nm, according to artificially varying $M_s$ (the saturation magnetization), $A_{ex}$, and $\gamma$ (the gyromagnetic ratio), using $M_s/M_{s,Py} = 1.0 \sim 2.0$, $(A_{ex}/A_{ex,Py})^{1/2} = 0.75 \sim 1.5$, and $\gamma/\gamma_{Py} = 0.5 \sim 1.75$, where $M_{s,Py} = 860$ emu/cm$^3$, $A_{ex,Py} = 1.3$ $\mu$erg/cm, $\gamma_{Py} = 2.8 \times 2\pi$ MHz/Oe for Py. The simulation results in Figs. 2(c) and 2(d) display a linear increase of $v_{cri}$ with $(A_{ex}/A_{ex,Py})^{1/2}$ with an equal slope for different values of $M_s$, and a linear increase with $\gamma/\gamma_{Py}$, respectively [the $v_{cri}$ dependence on $\alpha$ is also shown in Supplementary Fig. 1 [23]).

All the simulation results confirm an explicit analytical form of the critical velocity, $v_{cri} \simeq \eta\gamma\sqrt{A_{ex}}$ with the proportional constant $\eta = 1.66 \pm 0.18$. To check the universality of this constant for different magnetic materials, we also simulated the VC switching for various materials such as Ni, Fe, and Co [24]. The values of $\eta_{Fe} = 1.72 \pm 0.09$, $\eta_{Co} = 1.66 \pm 0.06$ and $\eta_{Ni} = 1.91 \pm 0.24$ for these materials are obtained using the corresponding critical velocities, $v_{cri}^{Fe} = 439 \pm 22$, $v_{cri}^{Co} = 507 \pm 19$, $v_{cri}^{Ni} = 320 \pm 40$ m/s, as shown in the inset of Fig. 2(a). All $\eta$ values are close to 1.66 and thus identical within the estimated errors. The $\eta = 1.66$ acts as an universal constant that relates the critical velocity for the VC switching and only the parameters of $A_{ex}$ and $\gamma$. Moreover, it is interesting that the equation for $v_{cri} \simeq 1.66\gamma\sqrt{A_{ex}}$ does not explicitly include $M_s$. This fact can be understood from the physical origin of the VC switching. As reported previously [13], the VC switching is induced by the gyrofield that originates from a dynamic deformation of **M** being concentrated around the moving VC. The resultant **M** deformation leads to a VC instability and eventually to the VC switching via pure dynamic processes of the nucleation and annihilation of a vortex-antivortex (V-AV) pair [5-7]. Owing to the dramatic **M** deformation employed over a ten nm length scale, the dominant contribution to the critical value of gyrofield is the short-range exchange interaction. Therefore, $A_{ex}$ is the dominant parameter in determining the critical gyrofield and in turn $\upsilon_{cri}$. Following the same argument, the long-range dipolar interaction gives a negligible contribution to $\upsilon_{cri}$, so that the dimensions and shape of a given dot, as well as $M_s$ do not affect the value of $\upsilon_{cri}$, as manifested itself in the equation $\upsilon_{cri} \simeq 1.66\gamma\sqrt{A_{ex}}$.

On the basis of the above results, $\upsilon_{cri}$ only depends on $A_{ex}$, but neither on the dot size nor the driving force parameters of $\omega_H$ and $H_0$. Consequently, we can construct phase diagrams of the VC switching criterion and switching times with respect to $\omega_H$ and $H_0$. Figure 3 shows the simulation results on the switching boundary diagram where the VC switching (gray) and non-switching (white) areas are differentiated by different symbols for the several different dimensions of the Py dots, e.g., [*R* (nm), *L* (nm)] = [150, 20], [150, 30], [300, 20], and [450, 20]. The switching event boundaries for all different dimensions of the dots form almost the same line, which reflects again the fact that $\upsilon_{cri}$ does not vary with the size of a given dot. Owing to the resonant excitation of the VC motions at $\omega_H = \omega_D$, the VC velocity can reach $\upsilon_{cri}$ through its gyrotropic motion even it is driven by only an extremely small field amplitude [5,9,10]. Resonant VC motions yield a deep valley on the switching boundary in the vicinity of $\omega_H = \omega_D$, and, thus, the threshold value of $H_0$ required for the VC switching has a minimum at $\omega_H = \omega_D$.

In addition, we theoretically derived more general explicit analytical equations representing the switching boundary, which distinguishes the event of the VC switching and non-switching based on the linearized Thiele's equation of motion. A detailed derivation is described in supplementary online material [23]. Here we chose $\mathbf{H}_{CCW}$ necessary for the resonant gyrotropic motion of the up-core and successive core switching to the down-core [9,10,20]. From the general solution of the VC equation of motion in the linear regime, the instantaneous velocity of the up-core motion as a function of time $t$ driven by $\mathbf{H}_{CCW}$ is found as $\upsilon(t) = \frac{1}{3}\gamma R H_0 \sqrt{\Omega^2 + F(\Omega,t)} / \sqrt{(1-\Omega)^2 + d^2\Omega^2}$, where $d = -D/|G|$, $G = 2\pi L M_s / \gamma$ is the gyrovector amplitude, $D < 0$ is the damping constant [18], and $\Omega = \omega_H / \omega_D$. The function $F(\Omega,t)$ represents the time variable velocity term of the transient VC motions and has a maximum at $t_m \approx \pi / |\omega_H - \omega_D|$ [23]. This $\upsilon(t)$ equation in turn allows us to analytically construct the boundary ($RH_0^C$) diagram of the VC switching in the $\Omega$-$RH_0$ plane by putting $\upsilon(t) = \upsilon_{cri}$ and $H_0 = H_0^C$, as written by $RH_0^C \approx (3\upsilon_{cri}/\gamma)\sqrt{(1-\Omega)^2 + d^2\Omega^2} / (\Omega + e^{-d\omega_d t_m})$. The numerical calculation of $RH_0^C(\Omega)$ using $\upsilon_{cri}^{Py} = 330 \pm 37$ m/s is displayed by the yellow-colored area in Fig. 3. On resonance ($\Omega = 1$), the minimum value of $RH_0^C$ is $3d\upsilon_{cri}/\gamma$, where $d = \alpha[1 + \ln(R/R_C)/2]$, $\alpha$ is the Gilbert damping parameter, $R_C$ is the VC radius [25], indicating that $H_0^C(\Omega = 1) \sim 3d\upsilon_{cri}/\gamma R$ is the lowest field strength required for the VC switching [26]. The analytical solution (yellow-colored area) is somewhat in discrepancy with the micromagnetic simulations (symbols), although they are similar in shape/trend. This discrepancy can be associated with the fact that the present simulations of the VC switching imply a nonlinearity of the VC gyrotropic motions when the motions are close to the switching event [27], whereas the above analytical equation of $RH_0^C$ assumes only a linear-regime vortex motion. However, this nonlinear effect could be compensated simply by multiplying a scaling factor $S_F = + 1.4$ to the above equation of $RH_0^C$. Recent experimental results by Curcic *et al.* [10] support well our predicted values of $H_0^C$. Their reported value of $H_0^C = 3.4$ Oe for a square dot of 500 nm width and 50 nm thickness[10] is in good agreement with our value, $H_0^C = 4.1$ Oe obtained from $S_F 3d\upsilon_{cri}/\gamma R$ on resonance, considering their system of $R = 250$ nm and $L = 50$ nm, and experimentally measured $\upsilon_{cri}^{Py} = 190$ m/s, although this value $\upsilon_{cri}^{Py} = 190$ m/s is much smaller than $330 \pm 37$ m/s obtained from our micromagnetic simulation [28].

Next, we construct a phase diagram of the switching time $T_S$, i.e., the time period necessary for the VC switching to complete itself from an initial equilibrium VC position. As already reported in Ref. [7], the VC switching occurs through the serial processes of the nucleation and annihilation of a V-AV pair around the initial VC position, following the maximum deformation of the entire **M** structure of the VC [7,13]. This process follows just after $\upsilon(t)$ reaches $\upsilon_{cri}$ through the gyrotropic motion [12]. Accordingly, $T_S$ consists of three different duration times as expressed by $T_S = \Delta t_g + \Delta t_d + \Delta t_{V-AV}$. The individual times can thus be determined by the three different processes: $\Delta t_g$ is the time period required for a VC to reach $\upsilon_{cri}$, $\Delta t_d$ is the time period between a time when $\upsilon(t)$ reaches $\upsilon_{cri}$ and a time when a V-AV pair is nucleated, and $\Delta t_{V-AV}$ is the time period during which the serial processes of the nucleation and annihilation of the V-AV pair occur until the VC switching is completed. To estimate each value of $\Delta t_g$, $\Delta t_d$, $\Delta t_{V-AV}$, the above indicated successive processes are differentiated according to the definition of each process, as shown in micromagnetic simulations for a Py dot of $R$ = 150 nm and $L$ = 20 nm (for details, see Ref. [23]). Then, each value of $\Delta t_g$, $\Delta t_d$, $\Delta t_{V-AV}$, and their sum $T_S$ are plotted as a function of $H_0$ in Fig. 4(a). For the low field strengths ($H_0 <$ 0.4 kOe), $\Delta t_g$ is on the order of a few ns and becomes increased with decreasing $H_0$. Note that $\Delta t_g$ is much longer than both $\Delta t_d$ (~50 ps) and $\Delta t_{V-AV}$ (~30 ps). For the higher field strengths ($H_0 >$ 0.5 kOe), however, $\Delta t_g$ reaches a few ps, being much less than $\Delta t_d$ and $\Delta t_{V-AV}$, according to the simulation results. However, the LLG equation of motion used in the simulations is invalid below 1 ps time scale, so that all the results below 1 ps (10 ps as an upper bound) are physically meaningless. Consequently, there is no doubt that for $H_0 <$ 0.4 kOe $T_S$ is determined by $\Delta t_g$, whereas for $H_0 >$ 0.5 kOe, $T_S$ is determined by $\Delta t_d$ in addition to $\Delta t_{V-AV}$, i.e. at least ~ 80 ps, only in the range of physically meaningful switching time. From an application point of view, such high field strengths are not necessary for the VC switching when applying the resonance frequency. In other words, in the range of $H_0 <$ 0.4 kOe, $\Delta t_g$ is sufficient to represent $T_S$. It is very interesting that $\Delta t_g$ can be calculated using the analytical equation of $\upsilon(t)$ along with the condition $\upsilon(t) = \upsilon_{\mathrm{cri}}$ at $t = \Delta t_{\mathrm{g}}$ and taking into account the nonlinear effects that can be also compensated by the scaling factor $S_{\mathrm{F}}$. For $\Omega = 1$ and $H_0 > H_0^C$ (non-switching for $H_0 < H_0^C$), $\Delta t_{\mathrm{g}}$ is analytically expressed as $\Delta t_{\mathrm{g}}(H_0) = -(d\omega_{\mathrm{D}})^{-1} \ln(1 - H_0^C / H_0)$, and this numerical solution is compared with the simulation result in Fig. 4(a).

For the case of $\Omega \neq 1$, $\Delta t_{\mathrm{g}}$ can also be numerically calculated, as shown in Fig. 4(b), e.g. for a Py dot of $R$= 150 nm and $L$= 20 nm. For $H_0 \gg H_0^C$, $\Delta t_{\mathrm{g}}$ is ~ 1 ps, and does not vary much with $\omega_{\mathrm{H}}$, but this time scale is physically meaningless, as already mentioned. In contrast, as $H_0$ decreases close to $H_0^C$, $\Delta t_{\mathrm{g}}$ markedly increases and varies considerably with $\omega_{\mathrm{H}}$. At $\Omega = 1$, it appears that the lower $H_0^C$, the longer $\Delta t_{\mathrm{g}}$ [29]. For faster VC switching, larger values of $H_0$ and $\omega_{\mathrm{H}} > \omega_{\mathrm{D}}$ are more effective. Since $\Delta t_{\mathrm{g}}$ of the order of 1 ps for $H_0 > 0.5$ kOe is not meaningful any more, and in such higher field region $\Delta t_{\mathrm{d}} + \Delta t_{\text{V-AV}}$ determines $T_{\mathrm{S}}$, only the region of $H_0 < 0.4$ kOe or $\Delta t_{\mathrm{g}} > 0.3$ ns in the phase diagram, shown in Fig. 4(b), would be technologically useful in the optimization of driving force parameters that reliably control the ultrafast VC switching.

In conclusion, the critical velocity, $\upsilon_{\mathrm{cri}} = 1.66\gamma\sqrt{A_{\mathrm{ex}}}$, of the VC gyrotropic motion has been found to serve as the universal criterion for vortex-core switching, e.g. $\upsilon_{\mathrm{cri}}^{\mathrm{Py}} = 330 \pm 37$ m/s. Based on this criterion, we derived phase diagrams of the vortex-core switching and the switching time with respect to the frequency and amplitude of circularly rotating field, which provide guidance for practical implementations of a dot array in the vortex states to information storage devices. These phase diagrams are also useful in the design of the dot dimensions and proper choice of materials as well as to optimize external driving forces for the reliable ultrafast VC switching at extremely low power consumption.

We thank H. Stoll for fruitful discussions. This work was supported by Creative Research Initiatives (ReC-SDSW) of MEST/KOSEF. P.F. was supported by the Director, Office of Science, Office of Basic Energy Sciences, Materials Sciences and Engineering Division, of the U.S. Department of Energy.

**References**

* To whom all correspondence should be addressed: sangkoog@snu.ac.kr

[1] A. Hubert, and R. Schäfer, *Magnetic Domains* (Springer-Verlag, Berlin, New York, Heidelberg, 1998).

[2] S.-K. Kim, J. B. Kortright, and S.-C. Shin, Appl. Phys. Lett. **78**, 2742 (2001); S.-K. Kim *et al.*, Appl. Phys. Lett. **86**, 052504 (2005).

[3] T. Shinjo *et al.*, Science **289**, 930 (2000); A. Wachowiak *et al.,* Science **298**, 577 (2002); J. Miltat and A. Thiaville, Science **298**, 555 (2002).

[4] R. P. Cowburn, Nature Mater. **6**, 255 (2007); J. Thomas, Nature Nanotech. **2**, 206 (2007).

[5] B. Van Waeyenberge *et al.*, Nature **444**, 461 (2006).

[6] R. Hertel, S. Gliga, M. Fähnle, and C. M. Schneider, Phys. Rev. Lett. **98**, 117201 (2007). Q. F. Xiao *et al.*, J. Appl. Phys. **102**, 103904 (2007).

[7] K.-S Lee, K. Y. Guslienko, J.-Y. Lee, and S.-K. Kim, Phys. Rev. B **76**, 174410 (2007).

[8] V. P. Kravchuk *et al.*, J. Appl. Phys. **102**, 043908 (2007).

[9] S.-K. Kim, K.-S. Lee, Y.-S. Yu, and Y.-S. Choi, Appl. Phys. Lett. **92**, 022509 (2008).

[10] M. Curcic *et al.*, Phys. Rev. Lett. **101**, 197204 (2008).

[11] K. Yamada *et al.*, Nature Mater. **6**, 269 (2007).

[12] S.-K. Kim *et al.*, Appl. Phys. Lett. **91**, 082506 (2007).

[13] K. Y. Guslienko, K.-S. Lee, and S.-K. Kim, Phys. Rev. Lett. **100**, 027203 (2008).

[14] B. Krüger *et al.*, Phys. Rev. B **76**, 224426 (2007); M. Bolte *et al.*, Phys. Rev. Lett. **100**, 176601 (2008).

[15] S.-K. Kim *et al.*, IEEE Trans. Magn. (in press).

[16] See http://math.nist.gov/oommf.

[17] L. D. Landau and E. M. Lifshitz, Phys. Z. Sowjet. **8**, 153 (1935); T. L. Gilbert, Phys. Rev. **100**, 1243 (1955).

[18] A. A. Thiele, Phys. Rev. Lett. **30**, 230 (1973); D. L. Huber, Phys. Rev. B **26**, 3758 (1982).

[19] A linearly oscillating field can be decomposed into the CCW and clockwise (CW) rotating fields with the equal $H_0$ and $\omega_\mathrm{H}$ values [9].

[20] K.-S. Lee and S.-K. Kim, Phys. Rev. B **78**, 014405 (2008).

[21] K. Y. Guslienko *et al.*, J. Appl. Phys. **91**, 8037 (2002).

[22] J. P. Park *et al.*, Phys. Rev. B **67**, 020403(R) (2003); S. B. Choe *et al.*, Science **304**, 420 (2004); K. Y. Guslienko *et al.*, Phys. Rev. Lett. **96**, 067205 (2006).

[23] See EPAPS Document No. XXX for detailed descriptions. For more information on EPAPS, see http://www.aip.org/pubservs/epaps.html.

[24] For Fe, Co, and Ni, $A_{ex}$ = 2.1, 3.0, and 0.9 $\mu$erg/cm, $M_s$ = 1700, 1400, 490 emu/cm$^3$ were used, respectively, with the same value of $\gamma = 2.8 \times 2\pi$ MHz/Oe. In literatures, different values for those material parameters are used. However, we found from the simulations that the value of $\eta$ is almost identical within the error range.

[25] K. Y. Guslienko, Appl. Phys. Lett **89**, 022510 (2006).

[26] For CW rotating fields, $RH_0^C$ has a constant value of $3\upsilon_{cri}/\gamma$, which indicates that the $RH_0^C$ value of the $\mathbf{H}_{CW}$ is much larger than that of $\mathbf{H}_{CCW}$.

[27] K.-S. Lee and S.-K. Kim, Appl. Phys. Lett **91**, 132511(2007); K. S. Buchanan *et al.*, Phys. Rev. Lett. **99**, 267201 (2007).

[28] In contrast to micromagnetic simulation results on an ideal system, experimental values of $\upsilon_{cri}$ may be affected by other factors such as measurement errors, and the imperfections (nonmagnetic defects, grain boundaries, edge roughness), and different material paramters of real samples.

[29] From micromagnetic simulation results for *R* = 150 nm (300 nm) and *L* = 20 nm, $H_0^C$ is

only 14 Oe (7 Oe) and $T_\mathrm{S}$ is 17 ns (33 ns).

FIG. 1 (color online). (a) Magnetic vortex structure with either upward or downward core **M** orientation. The rotation sense of the local in-plane **M** around the VC is CCW. (b) The orbital trajectories (left) and velocity-versus-time curves of the up (red line) and down (blue line) cores in a circular Py nanodot of $R$ = 150 nm and $L$ = 20 nm driven by the indicated $\mathbf{H}_{CCW}$ of $\omega_H / \omega_D$ = 1 and $H_0$ = 20 Oe. Also, the velocities of VC motions driven by $\mathbf{H}_{CW}$ and $\mathbf{H}_{Lin}$ are plotted for comparison. The horizontal line (right) denotes the value of $v_{cri}^{Py}$ = 330 m/s with an estimated error of ± 37 m/s (gray-color region)

FIG. 2 (color online). (a) Instantaneous VC velocity $v(t)$ in the Py dot of $R$ = 150 nm and $L$ = 20 nm for different values of $H_0$ and $\Omega = \omega_H / \omega_D$. The inset shows the VC velocities versus time for different dots made of Fe, Co, and Ni. (b), (c), and (d) show the dependences of $v_{cri}$ on the dimensions of the Py dot, the material parameters of $A_{ex}$, $M_s$, and $\gamma$, respectively. All the results were obtained from the micromagnetic simulations. The gray-colored regions with the lines indicate the value of $v_{cri}$ with its error range ± 37 m/s.

FIG. 3 (color online). The boundary diagram of the VC switching driven by the field $\mathbf{H}_{CCW}$ in the $\Omega - RH_0$ plane. Blue circle, green triangle, orange square and black diamond symbols correspond to the simulated results for [$R$ (nm), $L$ (nm)] = [150, 20], [150, 30], [300, 20], and [600, 20], respectively. Yellow (dark gray) colored area indicates the switching boundary obtained using the analytical equation of $RH_0^C$ shown in the text for Py dots with $R = 50 \sim 5000$ nm and $L = 20 \sim 80$ nm, based on the estimated value of $v_{cri}^{Py} = 330 \pm 37$ m/s. The light purple (gray) colored area is the result of the multiplication of $S_F = 1.4$ to the equation of $RH_0^C$. The right side (blue-colored) axis indicates the result corresponding to specific dimensions, [$R$ (nm), $L$ (nm)] = [150, 20].

FIG. 4 (color online). (a) Time periods for the individual indicated processes as a function of the amplitude $H_0$ of $\mathbf{H}_{CWW}$ at $\omega_H/2\pi = \omega_D/2\pi = 580$ MHz, obtained from micromagnetic simulations (symbols with lines) for a Py dot of $R = 150$ nm and $L = 20$ nm, being compared with the numerical calculation (line) of the analytical form of $\Delta t_g$. (b) Contour plot of the VC switching time in the $\Omega - H_0$ plane, obtained from the numerical calculation of $\Delta t_g$ for the Py dot with $R = 150$ nm and $L = 20$ nm. $\Delta t_g$ only in the region below the indicated white dashed line determines $T_S$.

**FIG.1**.

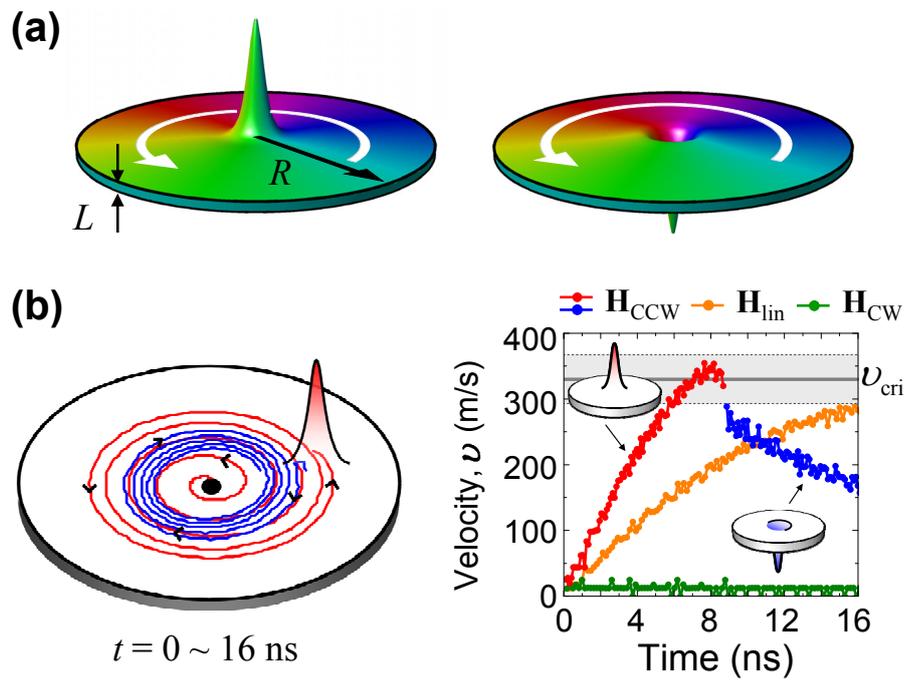

**FIG.2**.

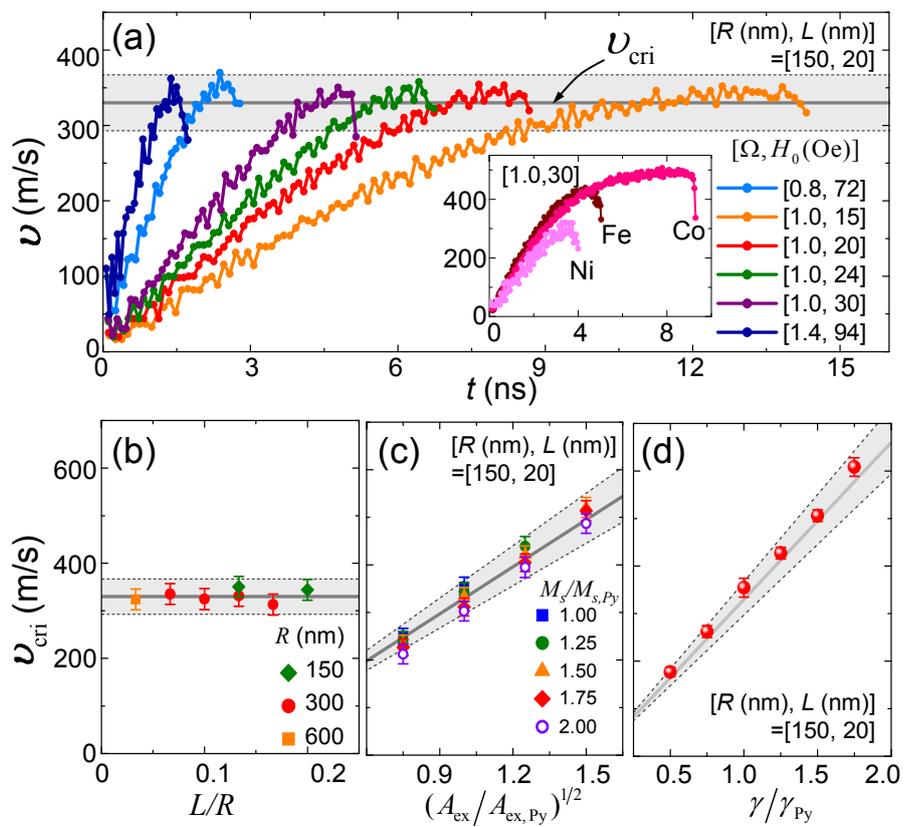

**FIG.3**.

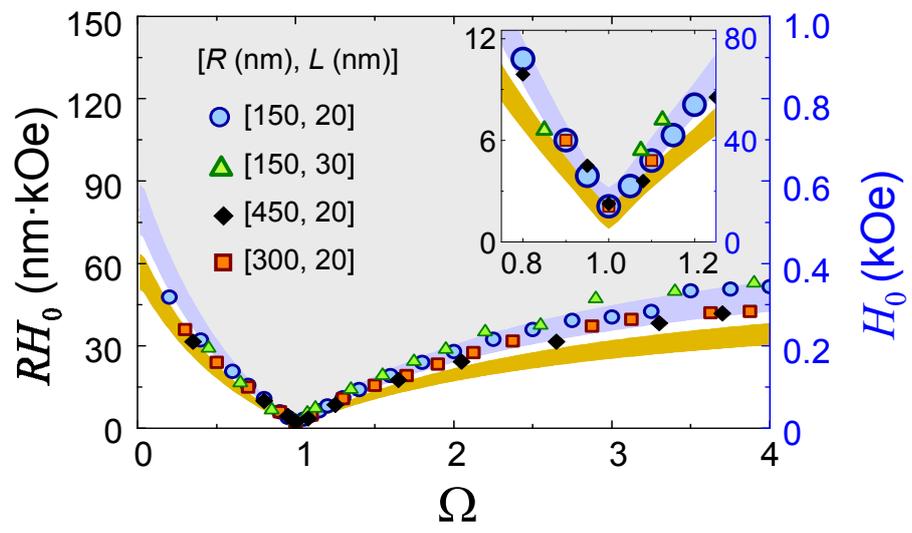

**FIG.4**.

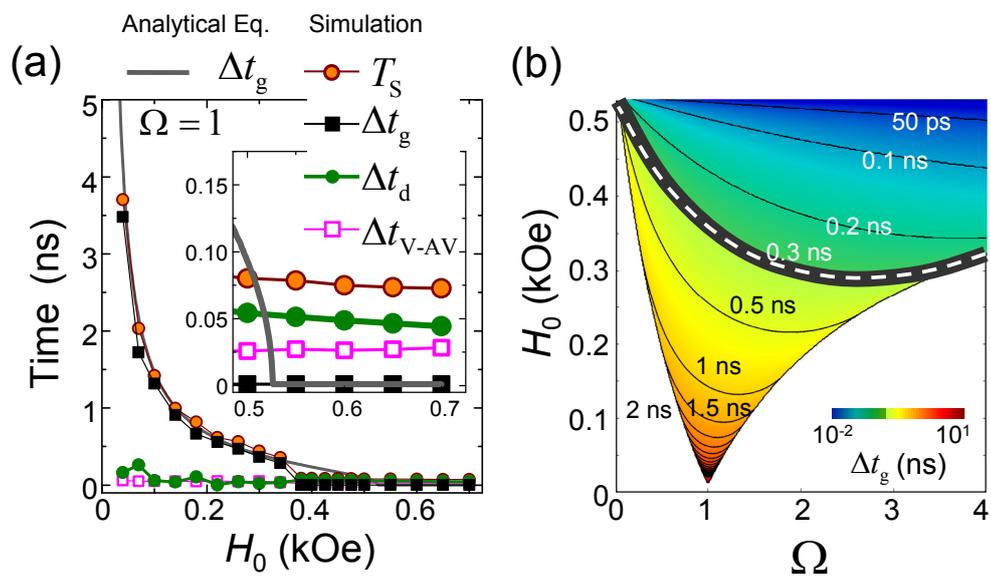

# Supplementary Documents

### A. Derivation of the instantaneous VC velocity, $\upsilon(t)$

To analytically derive the velocity of vortex-core motions in the linear regime, we derived the general solution of the vortex-core motion in both the earlier transient and later steady states [S1] based on the linearized Thiele's equation [S2]: $-\mathbf{G} \times \dot{\mathbf{X}} - \hat{D}\dot{\mathbf{X}} + \partial W(\mathbf{X},t)/\partial \mathbf{X} = 0$ with the vortex-core position $\mathbf{X} = (X, Y)$, the gyrovector $\mathbf{G} = -G\hat{\mathbf{z}}$, the damping tensor $\hat{D} = D\hat{I}$ (the identity matrix $\hat{I}$ and the damping constant $D < 0$), and the potential energy function $W(\mathbf{X},t)$ [S1,S3]. To solve the elementary rotating eigenmotions of a vortex core in a circular dot driven by the counter-clockwise (CCW)- and clockwise (CW)-circular rotating fields ($\mathbf{H}_{CCW}$ and $\mathbf{H}_{CW}$) [S4,S5], it is useful to utilize a complex variable $S \equiv X + iY$ for a vortex-core position $\mathbf{X} = (X, Y)$ at a time $t$, where the real and imaginary values indicate the $x$- and $y$- positions of the vortex core, respectively. For the case of applications of $\mathbf{H}_{CCW}$ for a selective switching from the up- to down-core, the general solution in the linear regime is given as $S = [S(0) - S_0]\exp(i\omega_D t)\exp(-d\omega_D t) + S_0 \exp(i\omega_H t)$ [S6] with $d = -D/|G|$ and $\omega_D = \kappa|G|/(G^2 + D^2)$ [S4,S5], where $\kappa$ is the stiffness coefficient of the potential energy $W(S, t)$, $S_0 = i\mu H_0/(\kappa - \omega_H G - i\omega_H D)$, and $S(0) = X_0 + iY_0$. $\mathbf{X}(0) = (X_0,$

$Y_0$) corresponds to the initial vortex-core position in an equilibrium state at $t = 0$. From the time derivative of this general solution of the vortex-core position **X**, the velocity of the up-core motion driven by $\mathbf{H}_{\mathrm{CCW}}$ is written as

$$\upsilon(t) = \frac{1}{3}\gamma R H_0 \frac{\sqrt{\Omega^2 + F(\Omega,t)}}{\sqrt{(1-\Omega)^2 + d^2\Omega^2}}$$

with the transient term of

$$F(\Omega,t) = \exp(-2d\omega_D t) - 2\Omega\exp(-d\omega_D t)\left[\cos\left((1-\Omega)\omega_D t\right) - d\left[\sin\left((1-\Omega)\omega_D t\right)\right]\right] \quad ,$$

where $\Omega = \omega_H/\omega_D$. For a sufficiently large value of $t$, the function $F(\Omega,t)$ converges to 0, i.e., the vortex motions arrive at certain pure steady states and the earlier transient states disappear.

**B. Entire switching time $T_S$ for complete vortex-core switching**

Ultrafast vortex-core switching in a soft magnetic nanodot is known to take place through the serial processes of the nucleation and annihilation of a vortex-antivortex pair around the initial vortex-core position [S7], following the maximum deformation of the entire magnetization structure of the initial vortex core [S7,S8]. This maximization of vortex-core deformation can be established just after the vortex-core velocity reaches its threshold velocity $v_{cri}$ through its characteristic gyrotropic motion [S9]. Thereby, the entire switching time $T_S$, necessary for the vortex-core switching to complete itself from an initial equilibrium vortex-core position, consists of three different duration times, which are determined by the three different individual processes described above, such that, $T_S = \Delta t_g + \Delta t_d + \Delta t_{V\text{-}AV}$. $\Delta t_g$ is the time period required for a vortex-core to reach $v_{cri}$ through the gyrotropic motion, $\Delta t_d$ is the time period during which a dynamic deformation of the entire magnetization of the initial vortex core is maximized before the nucleation process of a vortex-antivortex pair starts and after the initial vortex-core velocity reaches $v_{cri}$, and $\Delta t_{V\text{-}AV}$ is the time period during which the serial processes of the nucleation and annihilation of a vortex-antivortex pair take place until vortex-core switching is completed.

To obtain the entire switching time $T_S$ for complete vortex core switching as a function of the field amplitude and frequency of counter-clockwise circularly rotating magnetic fields, it is necessary to define each boundary between the individual successive processes. Supplementary Figure 2 shows the three different duration times, obtained from micromagnetic simulations for vortex-core switching from the up- to down-core in a Permalloy dot of $R = 150$ nm and $L = 20$ nm, driven by a counter-clockwise rotating field of $H_0 = 40$ Oe and $\omega_H/2\pi = \omega_D/2\pi = 580$ MHz. Each boundary between the successive two processes is indicated by the thick gray-colored vertical lines. The three different boundaries are denoted as "A", "B", and "C" in Suppl. Fig. 1 (the thick lines indicate the corresponding errors). The "A" represents when the instantaneous velocity $\upsilon(t)$ reaches $\upsilon_{cri}$, "B" indicates when the minimum value of the out-of plane magnetization component $m_z$, which represents the degree of deformation [S8], reaches $-0.9$ for the up- to down-core switching, and "C" denotes when the exchange energy reaches its maximum (this energy term is maximized when the vortex-antivortex pair is annihilated [S7]).

**References**

[S1] K.-S. Lee and S.-K. Kim, Appl. Phys. Lett. **91**, 132511 (2007).

[S2] A. A. Thiele, Phys. Rev. Lett. **30**, 230 (1973); D. L. Huber, Phys. Rev. B **26**, 3758

(1982).

[S3] K. Y. Guslienko, V. Novosad, Y. Otani, H. Shima, and K. Fukamichi, Phys. Rev. B **65**, 024414 (2002); K. Y. Guslienko, Appl. Phys. Lett. **89**, 022510 (2006).

[S4] K.-S. Lee and S.-K. Kim, Phys. Rev. B **78**, 014405 (2008).

[S5] S.-K. Kim, K.-S. Lee, Y.-S. Yu, and Y.-S. Choi, Appl. Phys. Lett. **92**, 022509 (2008).

[S6] K.-S. Lee and S.-K. Kim (unpublished).

[S7] K.-S Lee, K. Y. Guslienko, J.-Y. Lee, and S.-K. Kim, Phys. Rev. B **76**, 174410 (2007).

[S8] K. Y. Guslienko, K.-S. Lee, and S.-K. Kim, Phys. Rev. Lett. **100**, 027203 (2008).

[S9] S.-K. Kim, Y.-S. Choi, K.-S. Lee, K. Y. Guslienko, and D.-E. Jeong, Appl. Phys. Lett. **91**, 082506 (2007).

# Supplementary Figures

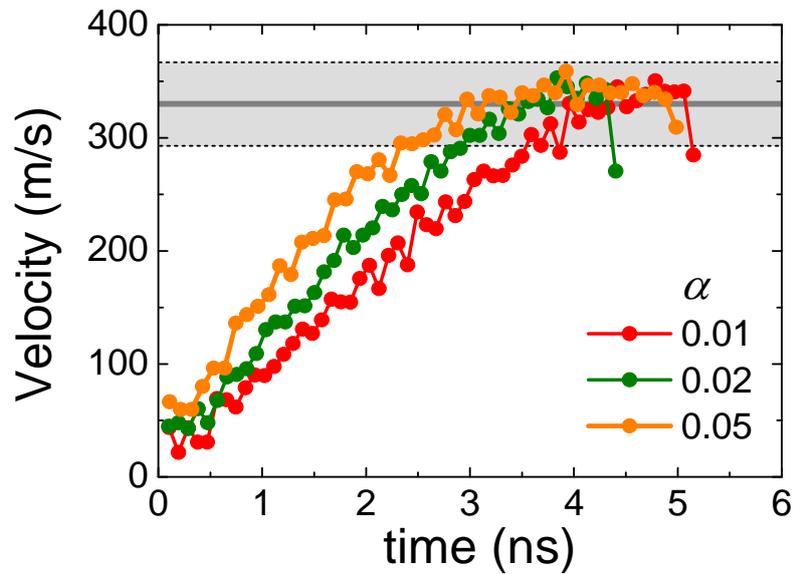

**SUPPL. FIG. 1.** (a) Instantaneous VC velocity $\upsilon(t)$ in the Py dot of $R = 150$ nm and $L = 20$ nm for different values of the damping constant $\alpha$ as noted. All the results were obtained from micromagnetic simulations. The gray-colored horizontal line and region indicate the value of $\upsilon_{cri} = 330$ m/s and its error range of $\pm 37$ m/s.

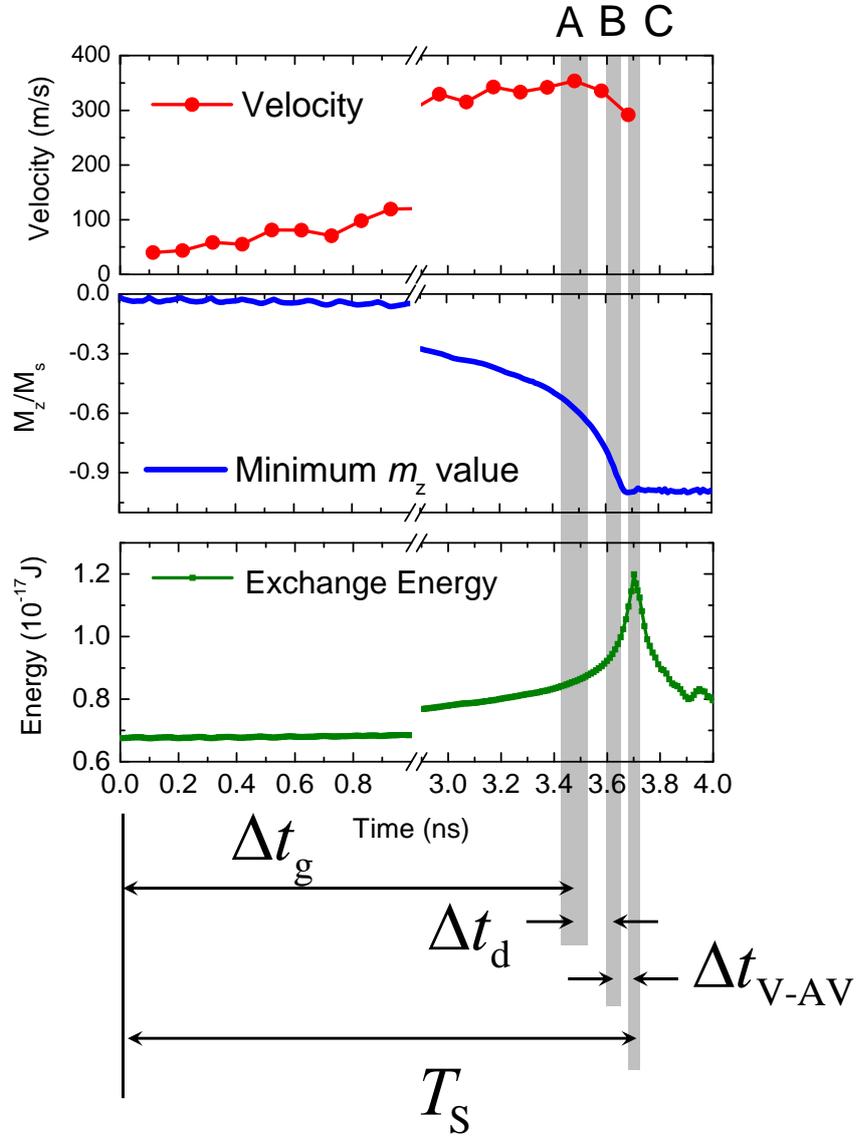

**SUPPL. FIG. 2.** Duration times necessary for the individual processes for complete vortex-core switching and boundaries between the individual successive processes. The results were obtained from micromagnetic numerical calculations for a Permalloy dot of $R = 150$ nm and $L = 20$ nm, and the CCW circularly rotating field with $H_0 = 40$ Oe and $\nu_H = \nu_D = 580$ MHz.